\documentclass[aps,twocolumn,showpacs,preprintnumbers,amsmath,amssymb,floatfix]{revtex4}

\usepackage{graphicx}
\usepackage{dcolumn}
\usepackage{bm}

\begin{document}

\title{Equilibrium roughening transition in a 1D modified sine-Gordon model}

\author{Sa\'ul Ares}
\email{saul@math.uc3m.es}
\affiliation{Grupo Interdisciplinar de Sistemas Complejos
(GISC) and
Departamento de Matem\'aticas,
Universidad Carlos III de Madrid, Avenida de la Universidad 30, 28911
Legan\'es, Madrid, Spain
}%

\author{Angel S\'anchez}
\homepage{http://gisc.uc3m.es/~anxo}
\affiliation{Grupo Interdisciplinar de Sistemas Complejos
(GISC) and
Departamento de Matem\'aticas,
Universidad Carlos III de Madrid, Avenida de la Universidad 30, 28911
Legan\'es, Madrid, Spain, and \\
Instituto de Biocomputaci\'on y F\'\i sica de Sistemas Complejos, 
Universidad de Zaragoza, 50009 Zaragoza, Spain
}%

\date{\today}

\begin{abstract}
We present a modified version of the one-dimensional sine-Gordon model that exhibits a
thermodynamic, roughening phase transition, in analogy with the 2D usual sine-Gordon
model. The model is suited to study the crystalline growth over an
impenetrable substrate and to describe the wetting transition of a liquid that
forms layers. We use the transfer integral technique to write
down the pseudo-Schr\"odinger
equation for the model, which allows to obtain some analytical insight,
and to compute numerically the free energy from
the exact transfer operator. We compare the results
with Monte Carlo simulations of the model, finding a perfect agreement between
both procedures. We thus establish that the model shows
a phase transition between a low
temperature flat phase with intriguing non trivial properties
and a high temperature rough one.  The fact that the model is one
dimensional and that it has a true phase transition makes it an ideal framework for
further studies of roughening phase transitions.
\end{abstract}


\maketitle

\section{Introduction}
\label{sec:int}

The two dimensional (2D)
ordered sine-Gordon model is today a fairly well understood problem (see, e.g.,
\cite{Plischke,Weeks,Beijeren,Barabasi,Krug,Pimpinelli}). However, the random
version of the model, where quenched disorder is introduced, is far less understood
and still subject of discussion. Since the {\em super-roughening}
transition (see Sec.\ \ref{sec:mod} below for a definition)
was introduced
in 1990 \cite{Toner}, there has been no theoretical agreement about its nature and
the properties of the super-rough phase (see \cite{Shapir} for a review, see \cite{us}
for more references). Large-scale
numerical simulations or exact optimization results
\cite{Riegercom,Marinari,Lancaster,Zeng,Rieger1,Coluzzi,us1,jjfises,Rieger2},
were not enough to solve the question, due to its highly demanding computational
character. To circumvent this problem, in \cite{super} we proposed a
modification of the one dimensional (1D) model, in order to have a less demanding
problem that could give us information about the super-rough phase. In the present
work, we proceed to a detailed characterization of the ordered version of the
model in \cite{super} and the roughening transition it presents. Having such 
a thorough analysis will not only serve as grounds for our results on 
super-roughening \cite{super} but will also be relevant from a much more 
general viewpoint, as a case study for 1D phase transitions and as an 
alternative way to obtain information about complicated problems in higher
dimensional systems.

Indeed, the first of the two goals above may seem questionable in view that
the subject of 1D thermodynamic phase transitions, defined as non-analyticities
of the free energy, has been for a long time excluded from
the attention of the community. This exclusion comes from the `public knowledge' 
that phase transitions can not occur in 1D systems with short range interactions. 
However, this general belief has risen
due to the misunderstanding of van Hove's theorem
\cite{vanHove,lieb} and abuse of Landau's \cite{landau} argument about
the entropic contribution
of domain walls to the free energy. In fact, there are many known examples of
this kind of transitions (see \cite{us4} for a comprehensive
study of the matter), although
most of them have been hidden using language tricks that have made us speak
about 1+1 dimensions. This has been the case, for instance, with a number of models of
the so called ``2D wetting'' that are in fact 1D in the mathematical sense,
as \cite{forgacs,derrida}, just to give a couple of examples.
In this sense, our work is yet another piece of detailed evidence about the
existence of 1D phase transitions. In addition, our model has immediate 
applications, such as 
crystalline growth over an impenetrable
substrate, or ``2D wetting'' favoring the
formation of layers of the condensed phase.
On the other hand, as mentioned in the previous paragraph,
it is clear that the same approach can be used for the study of other
2D problems, that may allow the formulation of a 1D counterpart as we present
in this paper for the sine-Gordon model.

With the above objectives in mind, the paper is organized as follows:
Section \ref{sec:mod} introduces our model by discussing in detail its
predecessors, the Burkhardt and the sine-Gordon ones. Subsequently, 
in Sec.\ \ref{sec:ana} we present the transfer
operator formalism and develop it into the pseudo-Schr\"odinger-equation
approximation that predicts a phase transition for the model.
We thus obtain analytical expressions for the magnitudes of interest
at low and high temperatures. Then in Sec.\ \ref{sec:num} we solve
numerically the transfer operator problem, showing the existence of the phase
transition and computing thermodynamical magnitudes such as the specific heat.
In Sec.\ \ref{sec:MC}
we present the results of Monte Carlo parallel tempering simulations of the
model, compare them with the preceding results and discuss the non-trivial
behavior in the flat phase of the model. Finally, in Sec.\ \ref{sec:conc} we
discuss the consequences of all this as well as further implications of our
results.

\section{Model definition}
\label{sec:mod}

In order to properly introduce and motivate our model, we find it convenient
to review in some detail the two previously proposed ones on which it is
based, namely 
the model introduced by Burkhardt in 1981 \cite{burkhardt} in the
context of wetting, and
the sine-Gordon model, a widely applicable model representative of a variety of
physical systems (see \cite{us} and references therein). Beginning with the
first one, the
Hamiltonian of Burkhardt's model is given by \cite{burkhardt}:
\begin{equation}
\label{eq:burk}
{\cal H}=\sum_{i=1}^{N}\Big\{J|h_{i+1}-h_i|-U(h_i)\Big\}.
\end{equation}
and defines a
continuous counterpart of the models proposed
by Chui and Weeks \cite{chui} and van Leeuwen and Hilhorst \cite{leeuwen}
in 1981.
We are interested in the version of the model with positive values of the
variables ($h_i\geq 0$). $U(h_i)$ is a function with a positive
constant value $U_0$ for
$h_i\leq R$ and zero for $h>R$.
The variables $h_i$ can be seen as heights over a substrate (located 
at $h=0$), defining all
together an interface.
This model is exactly solvable because its
transfer integral equation can be exactly mapped \cite{burkhardt}
to a Schr\"odinger equation, formally a
quantum square well problem in 1D with the square well potential at
the edge of the system. From quantum mechanics \cite{schiff} we know that this
potential has a bound state solution for a well deep enough. In Burkhardt's
statistical mechanical problem, the depth of the well of the resulting Schr\"odinger
equation depends on $\beta$, the inverse temperature. Hence, for low enough
temperatures, the quantum bound state maps to an interface trapped by the
potential, and therefore the interface is flat, in the sense that its width is
finite. Above the critical temperature of the model, the bound state disappears
and the interface depins from
the potential and its width diverges: it becomes rough.


The other pillar on which our model stands is 
the 1D sine-Gordon model, whose 
Hamiltonian is
\begin{equation}
\label{eq:sg}
{\cal H}=\sum_{i=1}^{N}\Big\{\frac{J}{2}(h_{i+1}-h_i)^2+
V_0[1-\cos(h_i)]\Big\},
\end{equation}
where now the values of the variables are unrestricted ($-\infty\leq h_i
\leq\infty $).
Its 2D version is very interesting because it can describe a number
of different physical problems \cite{us}
and because it
presents a {\em roughening} phase transition.
Again in the language of $h_i$ being the height of a surface, this transition
takes place
between a high temperature rough and a low
temperature flat phase (we will define more precisely below
what we understand by rough and
flat). In the rough phase, the roughness (also to be defined below,
but in the surface language can be thought of as the surface width)
of the
system scales as $\ln L$, the logarithm of the system size.
The roughening transition is modified by the addition of disorder to the
system:
when the cosine potential is changed adding a quenched disorder
$h^0_i$, making it $V_0[1-\cos(h_i-h^0_i)]$,
a {\em super-roughening} transition arises, characterized by the fact that the
low temperature phase is no longer flat. The super-roughening transition and
specially the low temperature phase are poorly understood.
One of the features that seem to be accepted about this {\em
super-rough} phase is that in it the roughness scales as $\ln ^2 L$, so in this
sense it would be even {\em rougher} than the high temperature rough phase
(hence the name super-rough).
Unfortunately,
the 1D version of the sine-Gordon model, much easier
to study analytically and numerically, is of no help to shed light on this
problem, as long as it has been rigorously proven \cite{us2} that it can not
have a true thermodynamic transition (although it does have an apparent one for
any finite size systems, even extremely large ones \cite{us3}). 

In order to better understand the phenomenology of the 2D version of the
model, in \cite{super} we introduced a new model containing the features of Burkhardt's  
and sine-Gordon models, in order to retain the most interesting
characteristics of both of them: the phase transition of the first one and the
periodic potential of the latter.
The rationale for this approach was that if we had a 1D model with such a phase
transition, we could consider its disordered version and check whether or not it
reproduces the features of 2D super-roughening. We indeed carried out this
program in \cite{super}, but a key question remained, namely whether or not the
basic, ordered, 1D model had a true thermodynamic phase transition or not. Only
if the answer of this question is positive will the approach in \cite{super} make
sense. Although our model is specifically designed to exhibit
this phase transition, and hence the transition itself would not be surprising,
we must prove hat the model behaves as expected: the fact that the model ingredients
suggest that it will indeed have a transition by no means warrants its existence.
In addition, the main purpose of this paper is the characterization of the
low temperature phase, which will show novel non trivial behavior as we will see below.

The Hamiltonian for
our model, which we
called the Burkhardt-sine-Gordon
model, (BsGM hereafter), is:
\begin{equation}
\label{eq:Bsg}
{\cal H}=\sum_{i=1}^{N}\Big\{\frac{J}{2}(h_{i+1}-h_i)^2+
V_0 \cdot V(h_i)\Big\},
\end{equation}
where
\begin{eqnarray}
\label{eq:BsgV}
V(x)= \left \{ \begin{array}{ll}
[1-\cos(x)]-\frac{1}{V_0}U(x) & \textrm{if $x\geq 0$},\\
\infty & \textrm{if $x<0$}.
\end{array} \right.
\end{eqnarray}


The choice for a quadratic coupling instead of an absolute value one as in \cite{burkhardt}
is to make our model as close as possible to the original sine-Gordon model.
In addition, the gaussian fluctuations of the heights that this coupling implies
can be simulated with higher efficiency using a heat bath algorithm \cite{toral,us3}.
We impose no explicit restriction over the values of $h_i$; it is the value of
the potential for $h_i<0$ what forces the variable to take only positive values.
This unlimited range of definition of the variable will be useful for the formal
operations we will perform. We see now that $U(x)$ can be seen as an attractive
potential binding the interface to the substrate. The cosine potential will
favor the growth forming layers at a distance $2\pi$ of each other. For
definiteness we choose the
parameters of the model to be $V_0=1$, $U_0=2$ and
$R=2\pi$. In that way, the substrate will attract the two first layers.
We have also performed simulations with different values of the parameters. In
that way, we can change the number of attracting layers, or the critical temperature,
but no new qualitative behavior is found.

In view of the above considerations, we expect the BsGM to have a phase transition
between a flat (or pinned) phase at low temperatures and a rough
(or depinned) one at high temperatures. That is exactly what we needed in order
to compare to the results on 2D super-roughening in disordered sine-Gordon 
models \cite{super}. However, in that previous work, we did little more than
providing plausibility arguments and simulation evidence for the existence of
such a transition, hence the necessity of the detailed, much more rigorous 
work presented here.  
To characterize the transition, we define the roughness or interface width, $w$,
as:
\begin{equation}
\label{eq:w}
w^2=\left\langle{\frac{1}{N}\sum_{i=1}^N\lbrack{h_i-\bar{h}}\rbrack^2}\right\rangle,
\end{equation}
where
\begin{equation} \label{medh}
\bar{h}\equiv\frac{1}{N}\sum_{i=1}^Nh_i \end{equation} is the mean
height, and averages $\langle\cdots\rangle$ are to be understood with respect
to a statistical weight given by the Gibbs factor, ${\rm e}^{-\mathcal{H}/T}$,
at equilibrium at a temperature $T$.
Then we say that an interface is flat when $w$
is finite and does not depend on the system size, $N$. In the rough
phase, the interface width grows with $N$ and diverges in the thermodynamic
limit, $w\to \infty$ as $N\to \infty$. 
Additionally, in the remainder of the paper we will look at 
other possible indicators of the transition, such as the 
free energy, the specific heat or the full correlation function.

\section{Analytical results.}
\label{sec:ana}

\subsection{Transfer integral technique.}
\label{sec:ana.tra}

The following discussion of the transfer integral (TI) technique follows that in
\cite{schneider} for the sine-Gordon model.
The classical canonical partition function of the BsGM [Eq.\ (\ref{eq:Bsg})]
can be written as:
\begin{equation}
\label{eq:part}
\mathcal{Z}_N(\beta)=\int_{-\infty}^{\infty}\textrm{d}h_1\int_{-\infty}^{\infty}
\textrm{d}h_2\cdots\int_{-\infty}^{\infty}\textrm{d}h_N
\textrm{e}^{-\beta \mathcal{H}},
\end{equation}
$\beta$ being the inverse temperature in units of the Boltzmann constant. Note
that we could have written the integrals in the range $[0,\infty)$, but our
definition of $V(h_i)$ makes this unnecessary.
In what follows, periodic boundary conditions
\begin{equation}
\label{eq:bc}
h_1=h_{N+1}
\end{equation}
are assumed, so that Eq.\ (\ref{eq:part}) can be replaced by
\begin{equation}
\label{eq:part.bc}
\mathcal{Z}_N(\beta)=\int_{-\infty}^{\infty}\textrm{d}h_1\cdots
\int_{-\infty}^{\infty}\textrm{d}h_{N+1}
\textrm{e}^{-\beta \mathcal{H}}\delta(h_1-h_{N+1}).
\end{equation}

To evaluate $\mathcal{Z}_N(\beta)$ we proceed as follows. The $\delta$ function
is represented as an expansion in a set of complete orthonormal functions
$\varphi_n(h)$:
\begin{equation}
\label{eq:delta}
\delta(h-h')=\sum_n \varphi_{n}^{*}(h)\varphi_n(h').
\end{equation}
The functions $\varphi_n$ are chosen to satisfy the the TI
equation:
\begin{equation}
\label{eq:TI}
\int_{-\infty}^{\infty}\textrm{d}h\exp[-\beta V_0 K(h,h')]\varphi_n(h)=
\exp(-\beta V_0 \epsilon_n)\varphi_n(h')
\end{equation}
where
\begin{equation}
\label{eq:TK}
K(h,h')=\frac{1}{2}\frac{J}{V_0}(h-h')^2+\frac{1}{2}[V(h)+V(h')],
\end{equation}
$\varphi_n$ is an eigenfunction of the TI equation with associated eigenvalue
$\exp(-\beta V_0 \epsilon_n)$, and
\begin{equation}
\label{eq:TO}
\mathcal{T}(\beta)=\exp[-\beta V_0 K(h,h')]
\end{equation}
is the transfer operator
of the model.
Using this we can rewrite the partition function:
\begin{align}
\label{eq:partfinal}
\mathcal{Z}_N(\beta)=&\sum_{n} \exp(-\beta V_0 \epsilon_n N)\int_{-\infty}^{\infty}
\textrm{d}h\varphi_n^*(h)\varphi_n(h)=\nonumber\\
=&\sum_{n} \exp(-\beta V_0 \epsilon_n N).
\end{align}
In the last step we have used the orthonormality of the $\varphi_n$. The
orthogonality and completeness of the eigenfunctions are guaranteed by the
Sturm-Liouville theory for Fredholm integral equations with a symmetric kernel
\cite{ch},
for which Eq.\ (\ref{eq:TK}) is an example.

Since the single site potential [Eq.\ (\ref{eq:BsgV})] is bounded from below, the
eigenspectrum is also bounded from below, and we denote the lowest eigenvalue by
$\epsilon_0$. This corresponds to $\exp(-\beta \epsilon_0)$,
the maximum
eigenvalue of the transfer operator (\ref{eq:TO}). In the thermodynamic limit,
the free energy per particle is then given by
\begin{equation}
\label{eq:free}
f=-k_BT\lim_{N\to\infty}\frac{1}{N}\ln\mathcal{Z}_N(\beta)=V_0 \epsilon_0.
\end{equation}

From this result, other thermodynamic properties can now be derived, i.e.,
internal energy per particle
\begin{equation}
\label{eq:energy}
e=\frac{1}{N}\langle \mathcal{H} \rangle=f-T\frac{\partial f}{\partial T}
\end{equation}
and specific heat at constant volume (length)
\begin{equation}
\label{eq:specific}
c_V=\frac{\partial e}{\partial T}=-T\frac{\partial^2f}{\partial^2T}.
\end{equation}

In \cite{schneider} it is also shown that in the thermodynamic limit
canonical averages are given by the expression
\begin{equation}
\label{eq:averages}
\langle g(h_i) \rangle = \int_{-\infty}^{\infty}|\varphi_0(h)|^2 g(h) \textrm{d}h,
\end{equation}
what means that $|\varphi_0(h)|^2$ can be understood as the probability density
for the variables $h_i$.

We have been unable to exactly evaluate the free energy (\ref{eq:free}) for the BsGM
[Eq. (\ref{eq:Bsg})]. As an alternative,
in this work we will proceed in two different ways:
through the pseudo-Schr\"odinger-equation approximation
associated to the TI equation
and computing numerically the
eigenvalues of the transfer operator.
In what follows we discuss the former approach, as well as the 
approximate analytical results that can be obtained from it. The
numerical study of the full transfer operator deserves a separate
treatment and is reported in Sec.\ \ref{sec:num}.


\subsection{Pseudo-Schr\"odinger-equation.}
\label{sec:ana.pse}

Defining
\begin{equation}
\label{eq:change}
\psi_n(h)=\exp(-\beta V_0 \frac{1}{2}V(h))\varphi_n(h)
\end{equation}
and using the identity
\begin{align}
\label{eq:identity}
\frac{1}{\sqrt{2\pi t}}\int_{-\infty}^{\infty}&\textrm{d}y
\left(\exp\left(-\frac{1}{2t}(x-y)^2\right)\right)f(y)=\nonumber\\
=&\exp\left(\frac{t}{2}\frac{\textrm{d}^2}{\textrm{d}x^2}\right)f(x),
\end{align}
the TI equation (\ref{eq:TI}) may be rewritten in the form
\begin{align}
\label{eq:expdif}
&\exp\left(\frac{1}{2\beta}\frac{1}{\sqrt{V_0J}}\sqrt{\frac{V_0}{J}}\frac{\textrm{d}^2}{\textrm{d}h^2}\right)
\psi_n(h)=\nonumber\\
&\exp\left[-\beta\sqrt{V_0J}\sqrt{\frac{V_0}{J}}(\epsilon_n-V_{eq}-V(h))\right]\psi_n(h)
\end{align}
where
\begin{equation}
\label{eq:C0}
V_{eq}=\frac{1}{2V_0\beta}\ln\left(\frac{\beta J}{2\pi}\right).
\end{equation}
In the limit of strong coupling ($J/V_0 \to \infty$
as we keep $V_0J$ constant, see \cite{schneider} for
details; it can also be understood as a continuum limit if we include the
lattice constant as an explicit parameter of the model)
this equation can be expanded to obtain, to first order in $V_0/J$:
\begin{equation}
\label{eq:schro}
\left[-\frac{1}{2\beta^2 V_0J}\frac{\textrm{d}^2}{\textrm{d}h^2}+V(h)\right]\psi_n(h)
=(\epsilon_n-V_{eq})\psi_n(h).
\end{equation}
This is the Schr\"odinger equation for a square well with a superimposed cosine
potential. The square well alone is exactly solvable, and for low values of
$V_0$ the cosine term can be treated with perturbation theory.
Eq.\ (\ref{eq:schro}) is already an approximation for our model;
from Eq.\ (\ref{eq:expdif}) we can see that
it is valid when $\sqrt{V_0/J} \ll \beta$ and $\sqrt{V_0/J} \ll 1/\beta$ .
If we take Boltzmann's constant as unity,
the predictions of this equation are expected to hold quantitatively only in the
the temperature region $\sqrt{V_0/J} \ll T \ll 1/\sqrt{V_0/J}$. For this
interval to make sense, $\sqrt{V_0/J}$ has to be small enough. For instance,
for the values of the parameters we use ($J=1, V_0=1$), there is no temperature
where the pseudo-Schr\"odinger equation is quantitatively accurate.
However, the qualitative picture this equation yields is completely valid and
describes correctly the phenomenology of the model. In the quantum mechanical
problem, for some values of the parameters of the model we have a bound state,
that disappears as we change the parameters. In our statistical mechanical
problem, fixing all the parameters except the temperature will give us a
thermodynamical phase transition between a flat phase at low temperatures,
pinned by the square well potential, and a rough phase at high temperatures,
where the interface has detached itself from the substrate's attraction. This
is the same scenario Burkhardt found in \cite{burkhardt}; the change of the
absolute value coupling for the quadratic one and the addition of the cosine
potential modify the quantitative aspects of the phase transition, but not the
qualitative ones. Of course, these new features in our model will give rise to new
phenomena in the flat phase's behavior. Anyway, if we make a further rough
approximation and dismiss the sinusoidal part of the potential in
Eq.\ (\ref{eq:schro}), we are left with exactly the Schr\"odinger equation of
a semi-infinite square well. From elementary quantum mechanics \cite{schiff}
(see also \cite{burkhardt} for the application to Burkhardt's model), we
know that the spectrum of this equation presents a continuum of scattering
states. For appropriate values of the parameters (that in the statistical mechanical
problem means $T<T_c$) there are one or more bound states. As $T\to T_c^-$, the
gap between the strongest bound state and the first scattering state varies as:
\begin{equation}
\label{eq:gap}
\Delta\epsilon\propto (T_c-T)^2.
\end{equation}


The quadratic temperature dependence of the gap in Eq.\ (\ref{eq:gap}) is
responsible for the finite jump in the specific heat of the model.
We will find this in the computation of
the specific heat both from the numerical transfer operator and from
Monte Carlo simulations. In Figure \ref{fig:theo} we
show the gap between the two first eigenvalues computed from the
exact numerical transfer operator; the quadratic behavior predicted in
Eq.\ (\ref{eq:gap}) is evident as $T\to Tc^-$.

For the rest of this work, without loss of generality,
we will take the coupling constant $J=1$. We can do this because the effect
of changing $J$ can be taken into account rescaling $V_0$, $U_0$ and the
temperature
(and also the time scale, but in this work we will deal only with equilibrium
properties).

\subsection{Low and high temperature approximations}

For low enough temperatures, it is a good approximation to suppose that
all the heights fall inside the square well
\noindent
\begin{figure}
\vspace*{2mm}
\includegraphics[width=8.0cm]{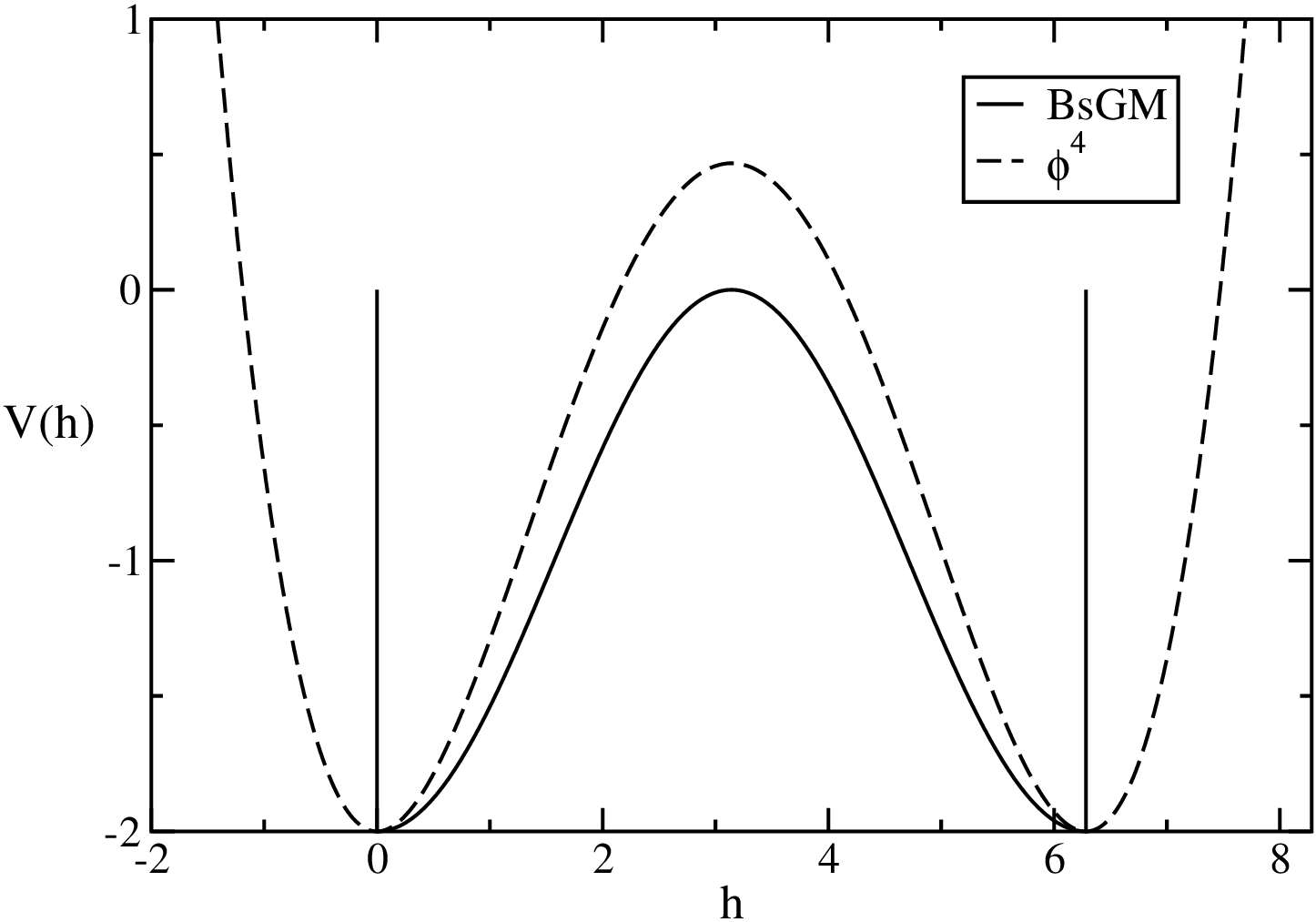}
\caption{\label{fig:phi4}Approximation of the potential inside the square
well by a $\phi^4$ potential. The continuous line is the BsGM potential between
$0$ and $2\pi$. The dashed line is the $\phi^4$ potential we use to approximate it.}
\end{figure}
potential. For a value of the width of the well of $R=2\pi$, inside the well
there exist two minima of the cosine potential. In that case it is
reasonable to approximate the potential by a $\phi^4$ one, see Fig.\
\ref{fig:phi4}. The good features of this choice are that the $\phi^4$ potential
reproduces the two potential minima and that it bounds the system to them, as it
grows to $\infty$ as $h \to \pm \infty$. Note that if we restrict ourselves to
only one minima of the
cosine potential, a parabolic potential will be enough to reproduce the leading
term. To mimic the potential in our problem, this $\phi^4$ potential has the form (for
$V_0=1$ and $U_0=2$):
\begin{equation}
\label{eq:phi4}
V_{\phi^4}(h)=\frac{(h-\pi)^4}{4\pi^2}-\frac{(h-\pi)^2}{2}+\frac{\pi^2}{4}-2
\end{equation}
In \cite{schneider} we find values for some thermodynamic properties of a
low temperature expansion of the $\phi^4$ model. Thus, we have for the internal
energy:
\begin{equation}
\label{eq:e_phi4}
e=\frac{T}{2}+\frac{36T^2}{15\cdot2^3\pi^2},
\end{equation}
and for the specific heat:
\begin{equation}
\label{eq:sh_phi4}
c_V=\frac{1}{2}+\frac{72T}{15\cdot2^3\pi^2}
\end{equation}
We will see that, at low temperatures, the system chooses to be in one single
potential minimum of the two displayed in Fig.\ \ref{fig:phi4}. In fact, this
assumption is implicit in the calculation that lead to Eqs.\ (\ref{eq:e_phi4})
and (\ref{eq:sh_phi4}) (see \cite{schneider} for details). This calculation
approximates the $\phi^4$ potential by a parabolic one ($V(h)=V_0h^2$), and then
introduces the higher order corrections.

The same procedure can be used with the sine-Gordon model instead of the $\phi^4$
model. It also seems a reasonable choice to approximate the $T\to 0$ regime
using this potential. In the end, as both models have the same leading term,
the differences between them will be small. We will compare the expressions arising
from both of them with the results of our simulations, and find that both of them
describe remarkably well physical magnitudes when $T\to 0$. From \cite{schneider},
we have the following expressions for the low temperature sine-Gordon model:
\begin{equation}
\label{eq:e_sG}
e=\frac{T}{2}+2\left[\left(\frac{T}{8}\right)^2+\left(\frac{T}{8}\right)^3
+\ldots\right]
\end{equation}
\begin{equation}
\label{eq:sh_sG}
c_V=\frac{1}{2}+2\left[\frac{2T}{8^2}+\frac{3T^2}{8^3}+\ldots\right]
\end{equation}


Both these approximations suppose the system is trapped in a single well of the
potential, and it can be seen that this implies that the system is in a flat
phase \cite{us3}. So agreement with these results is a signal of a flat phase.

Restricting ourselves to the lowest order approximation for vanishing temperatures,
that is, a flat system trapped in a single parabolic potential, it is
straightforward to calculate the roughness and correlation functions, as was done
in \cite{us3}. The parameter of the parabolic potential has to be
$V_0+U_0$. For the roughness we obtain:
\begin{equation}
\label{eq:ruglow}
w^2(T)=\frac{T}{\sqrt{(2+V_0+U_0)^2-4}}.
\end{equation}
We define the height-difference correlation function as:
\begin{equation}
\label{eq:corr}
C(r)=\left \langle{\frac{1}{N}\sum_j\lbrack{h_j-h_{j+r}}\rbrack^2}\right
\rangle.
\end{equation}
It can be shown that the parabolic potential approximation yields for it:
\begin{equation}
\label{eq:corrlow}
C(r)=\frac{2T}{\sqrt{(2+V_0+U_0)^2-4}}(1-C_c(r))
\end{equation}
where
\begin{equation}
\label{eq:ccr}
C_c(r)=\left(\left(1+\frac{V_0+U_0}{2}\right)\left[1-\sqrt{1-\left(\frac{2}{2+V_0+U_0}\right)^2}\right]\right)^r
\end{equation}
In the asymptotic limit $r\to \infty$, $C_c(r)\to 0$, and we have that
$C(\infty)=2w^2$.

In the high temperature phase, the potential effectively vanishes and we are
left with the quadratic coupling alone: this is the Edwards-Wilkinson model
\cite{EW}. The predictions for the internal energy ($e=T/2+\text{constant}$) and the specific
heat ($c_V=1/2$) are expected to hold in the rough phase of our model. However,
the prediction for the interface width is not so accurate: the existence in our
model of an impenetrable substrate changes the statistics of the rough
interface.

\section{Numerical transfer operator results}
\label{sec:num}


The eigenvalue problem in Eq.\ (\ref{eq:TI}) can be solved discretizing the
transfer operator in Eq.\ (\ref{eq:TO}) and evaluating numerically the
eigenvalues of the resulting matrix (see \cite{dp,theo,schneider};
see \cite{thierry} for a detailed account). The relevant parameters of the
discretization of the operator are $\Delta h$, the discretization length, and
$M$, the number of points considered, that is, the size of the matrix. From them
we obtain immediately the interval where the discretized variable takes values,
$[0,h_{max}]$, where $h_{max}=(M-1) \cdot \Delta h$. The two sources of error of
this numerical procedure are the discretization of the real variable $h$ and the
cutoff of the variable range at $h_{max}$. In the limit $\Delta h \to 0$
and $M \Delta h \to \infty$ (that is, $h_{max} \to \infty$),
this numerical approach is exact.


A thermodynamic phase transition takes place when there is a
non-analyticity in the free energy. We have seen in Eq.\ (\ref{eq:free})
that in the
thermodynamic limit the free energy is determined by the largest
eigenvalue of the transfer matrix. As discussed below Eq.\ (\ref{eq:gap}), the
vanishing of the gap between the largest two eigenvalues leads to a
singularity. To find the point of a phase transition, we have to find a
minimum of the gap and show that the minimum goes to zero as we increase $M$.

\noindent
\begin{figure}
\vspace*{2mm}
\includegraphics[width=8.0cm]{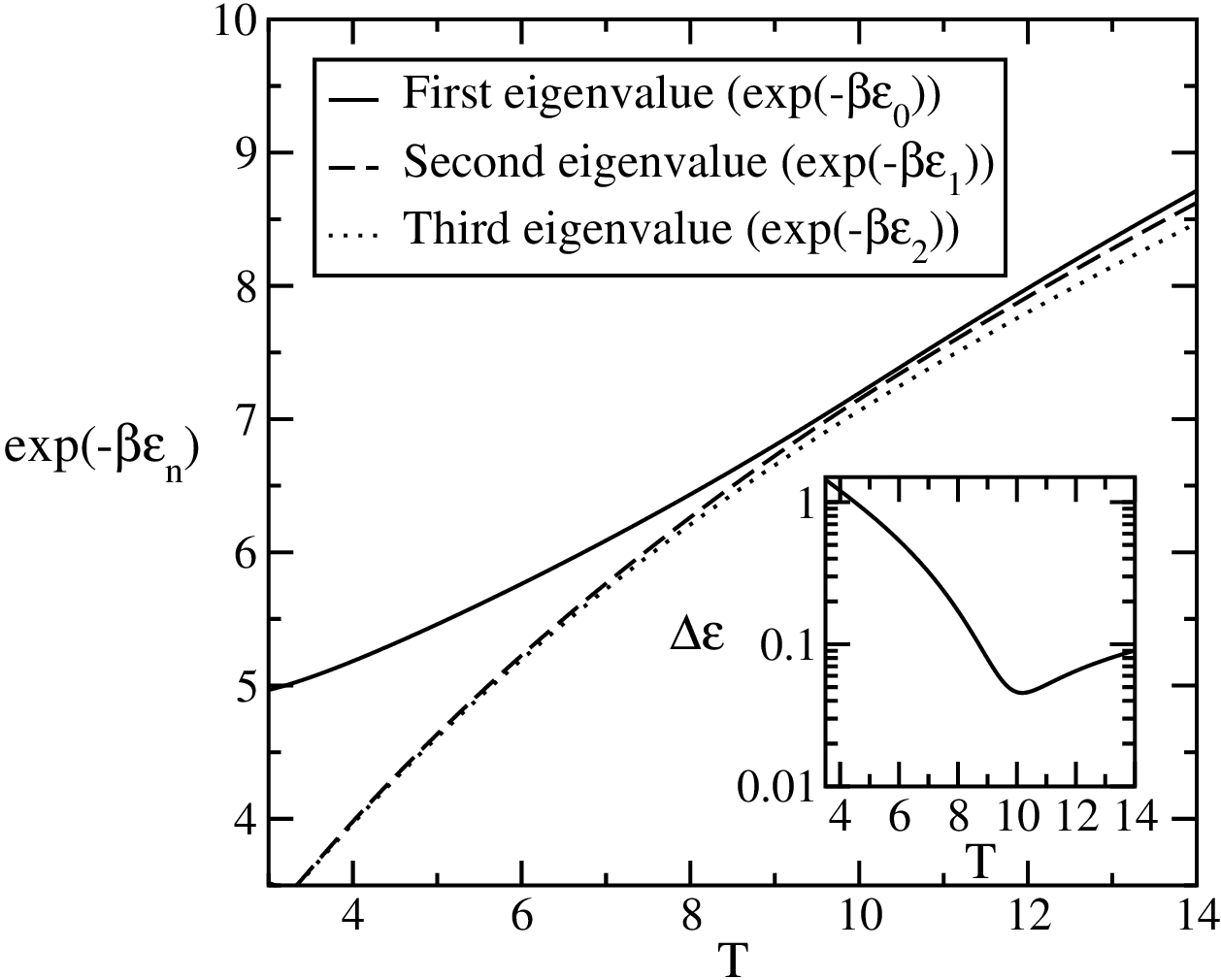}
\caption{\label{fig:eigs}Three first eigenvalues for $M=4096$ and $\Delta
h=1/32$. Inset: Difference $\Delta \epsilon=\exp(-\beta \epsilon_0)-
\exp(-\beta \epsilon_1)$ vs. $T$. The minimum gives the temperature of the phase
transition.}
\end{figure}

\begin{figure}
\vspace*{2mm}
\includegraphics[width=8.0cm]{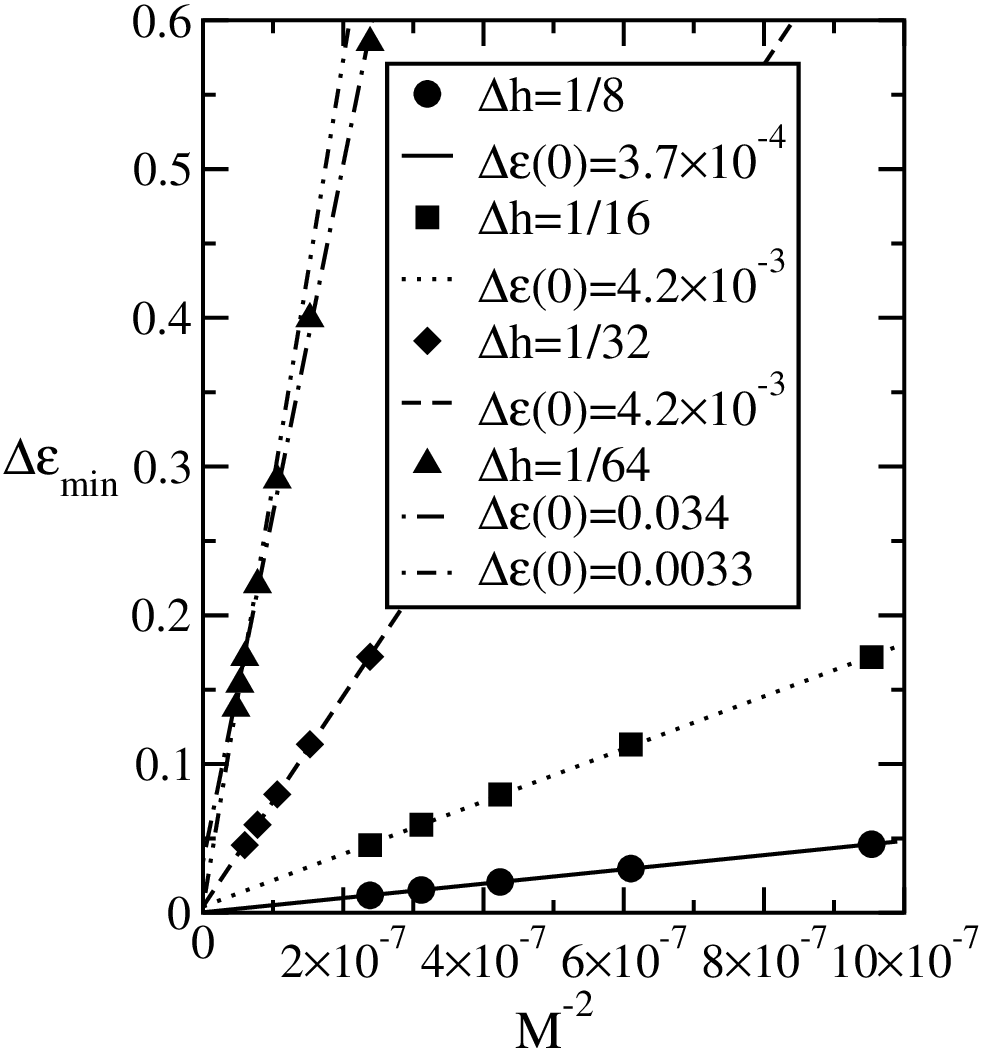}
\caption{\label{fig:gap}Minimum value of the gap for different discretization
values and matrix sizes, as indicated in the plot.}
\end{figure}

\begin{figure}
\vspace*{2mm}
\includegraphics[width=8.0cm]{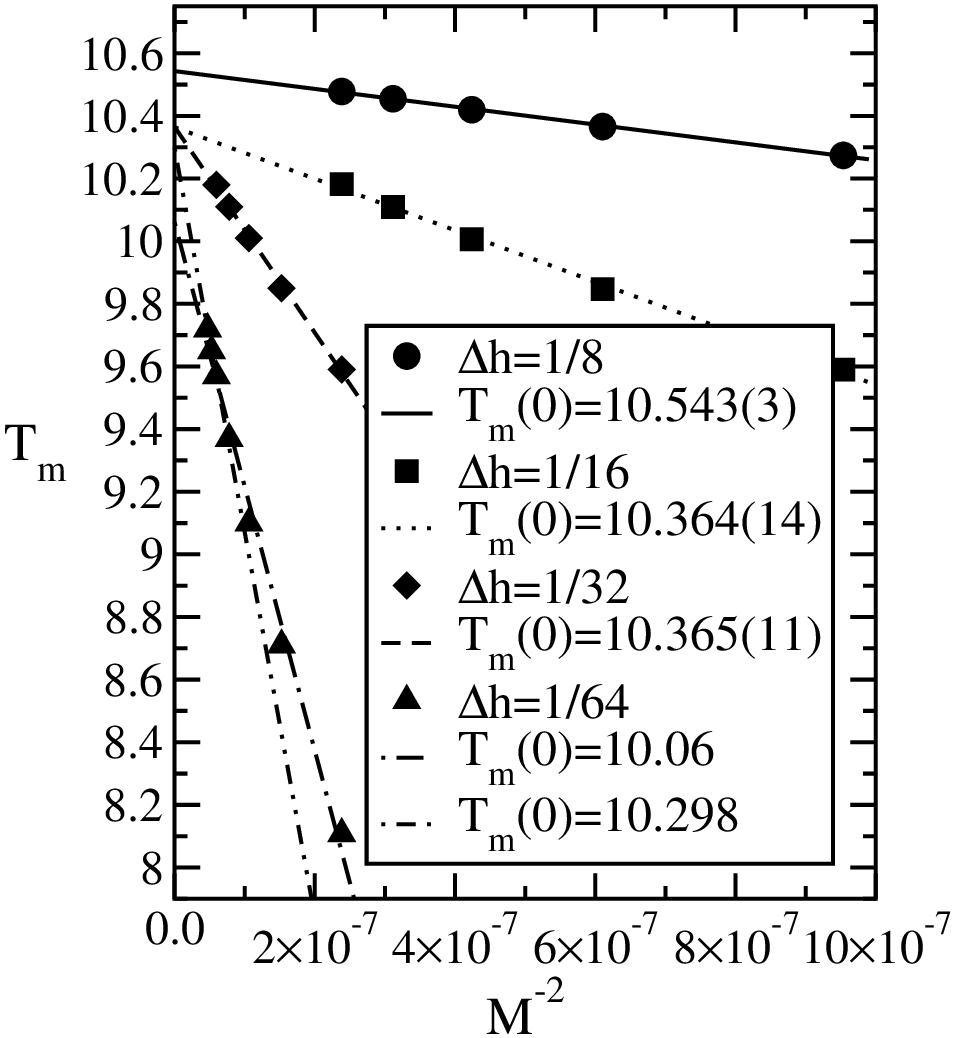}
\caption{\label{fig:Tc}Critical temperature for different discretization values
and matrix sizes as indicated in the plot.}
\end{figure}

In Figure \ref{fig:eigs} we show the first three eigenvalues of the
discretized transfer operator with our standard set 
of parameters, $V_0=1$, $U_0=2$ and $R=2\pi$. We clearly
see that the first two eigenvalues become very close near $T\approx 10$.
In the inset we show the minimum of $\Delta
\epsilon$ that indicates the temperature of the candidate 
transition. The slope of
$\epsilon_0$
does not change discontinuously at $T_m$, the temperature of the minimum,
so the transition will be continuous and not first order.
\noindent
\begin{figure}
\vspace*{2mm}
\includegraphics[width=6.0cm]{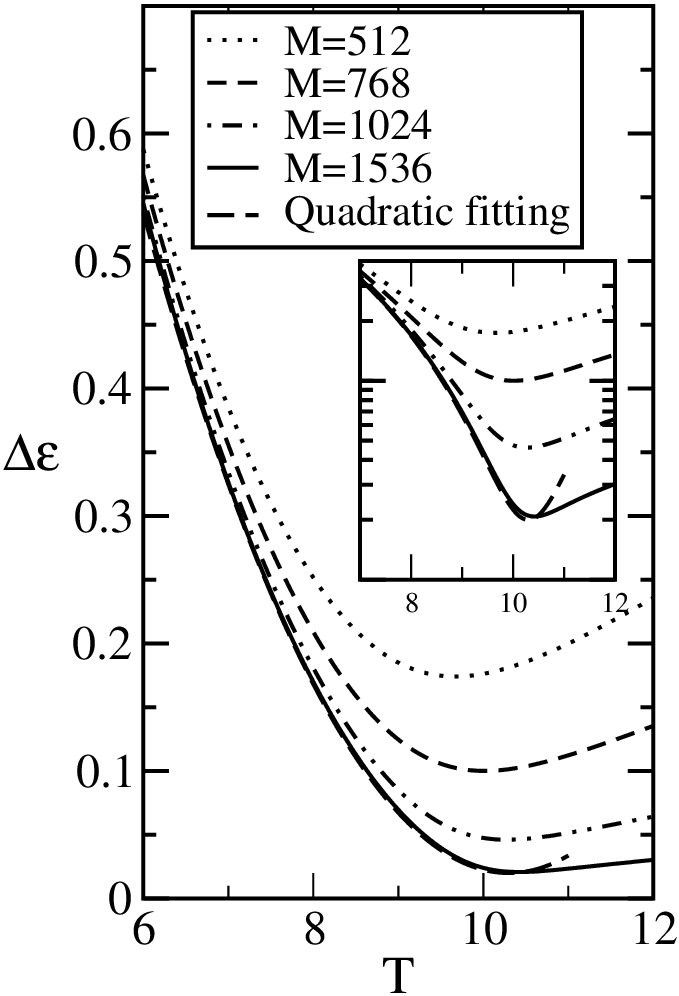}
\caption{\label{fig:theo}$\Delta \epsilon$ for different matrix sizes as
indicated in the plot. The
discretization is $\Delta h=1/8$. Inset: the same figure with the
$\Delta \epsilon$ axis in logarithmic scale. We see that as $M$ becomes greater,
$\Delta \epsilon$ goes quadratically to its minimum as $T\to T_c^-$,
as shown for $M=1536$ using a quadratic fit. This is exactly
the prediction of Eq.\ (\ref{eq:gap}).}
\end{figure}
In Figure \ref{fig:theo} we show the gap between the two first eigenvalues for a
range of matrix sizes, keeping $\Delta h$ fixed. We see that as $M$ 
increases, the minimum value of the gap becomes closer to zero. In Figures
\ref{fig:gap} and \ref{fig:Tc} we perform a finite-size scaling to check that
the minimum of the gap, $\Delta \epsilon_{min}$, goes to zero, and how the
different temperatures for the minimum go to the critical temperature, $T_c$.
We
\noindent
\noindent
see, as observed in \cite{theo} for a different model, that both $\Delta
\epsilon_{min}$ and $T_m$ scale with $M^{-2}$ when we change $M$ keeping $\Delta
h$ fixed. Of course, this scaling is supposed to improve for greater
matrix sizes, and this aspect is important specially for small $\Delta h$. In
Figure \ref{fig:gap} we see how as $M$ increases $\Delta \epsilon_{min}$ goes to
zero. It may seem contradictory that as we take a better (smaller) $\Delta h$,
the convergence to zero becomes worse. The explanation comes from the fact that,
as we use a smaller $\Delta h$, we need a bigger $M$ to get a correct scaling.
However, memory limitations of our computers sets a limit to the values of $M$ we
can use: we can not go much further of $M=4096$ in a reasonable time. So, to get
a better estimation of $\Delta \epsilon_{min}$, we use only the points with the
best scaling. That is what we do for $\Delta h=1/64$,
where using only the two points of greater $M$ we see that the
asymptotic value is corrected in one order of magnitude. We can then safely
expect $\Delta \epsilon_{min} \to 0$ as $M^{-2} \to 0$ and $\Delta h \to 0$.
This means that in fact we have a true thermodynamic phase transition, as
predicted by the pseudo-Schr\"odinger approximation. The critical temperature
$T_c$ can be inferred from the data in Figure \ref{fig:Tc}. The data coming from
the smallest values of $\Delta h$ are supposed to be the best one, and again we
have used only the last two values for $\Delta h=1/64$ to correct the effects of
lack of scaling for low $M$. With the data in the figure, we can estimate the
critical temperature as $T_c=10.3$ in our units.

\begin{figure}
\vspace*{2mm}
\includegraphics[width=8.0cm]{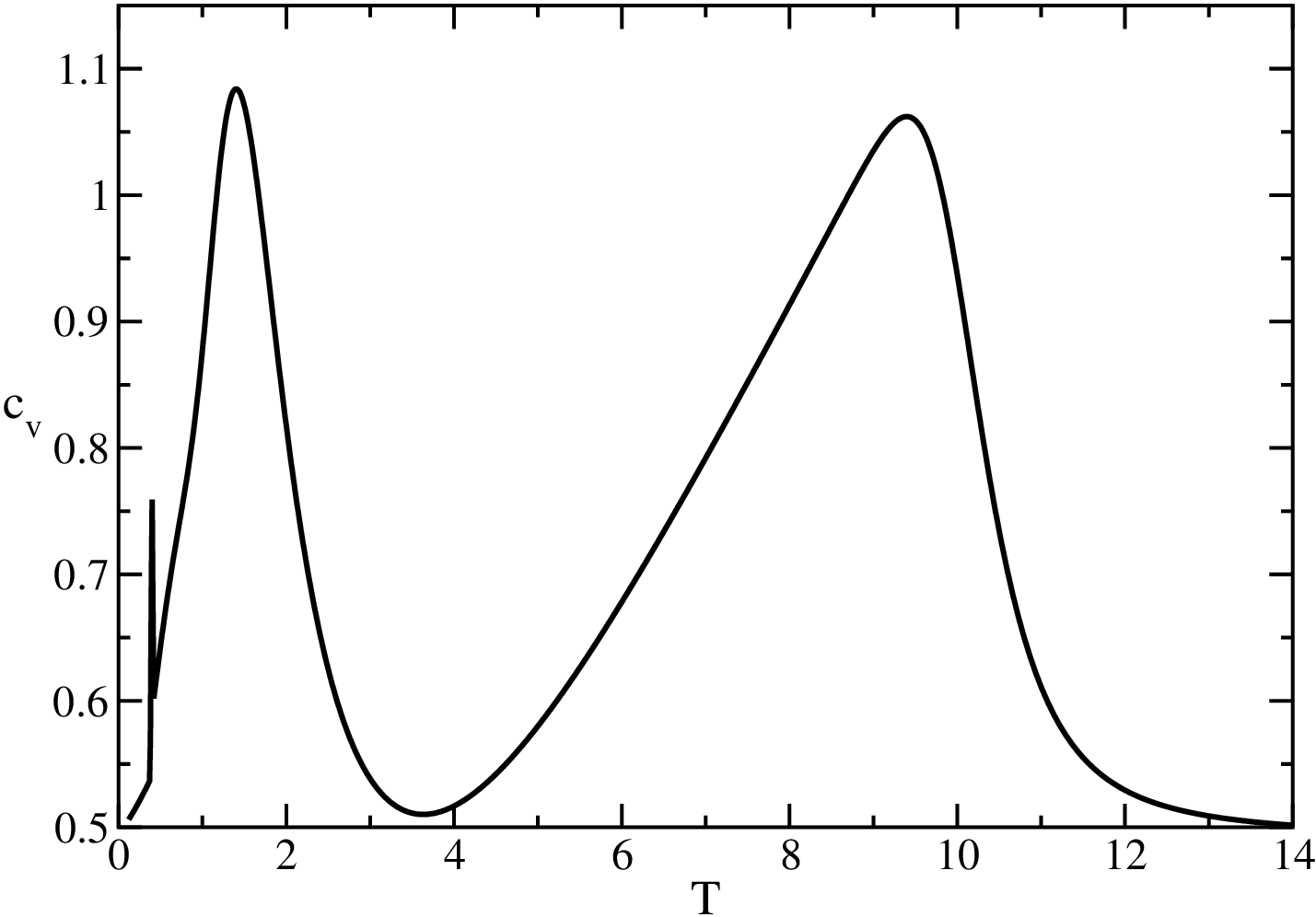}
\caption{\label{fig:shtrans}Specific heat as a function of temperature
obtained from the discretized transfer
operator for $\Delta h=1/32$ and $M=4096$.}
\end{figure}

\begin{figure}
\vspace*{2mm}
\includegraphics[width=8.0cm]{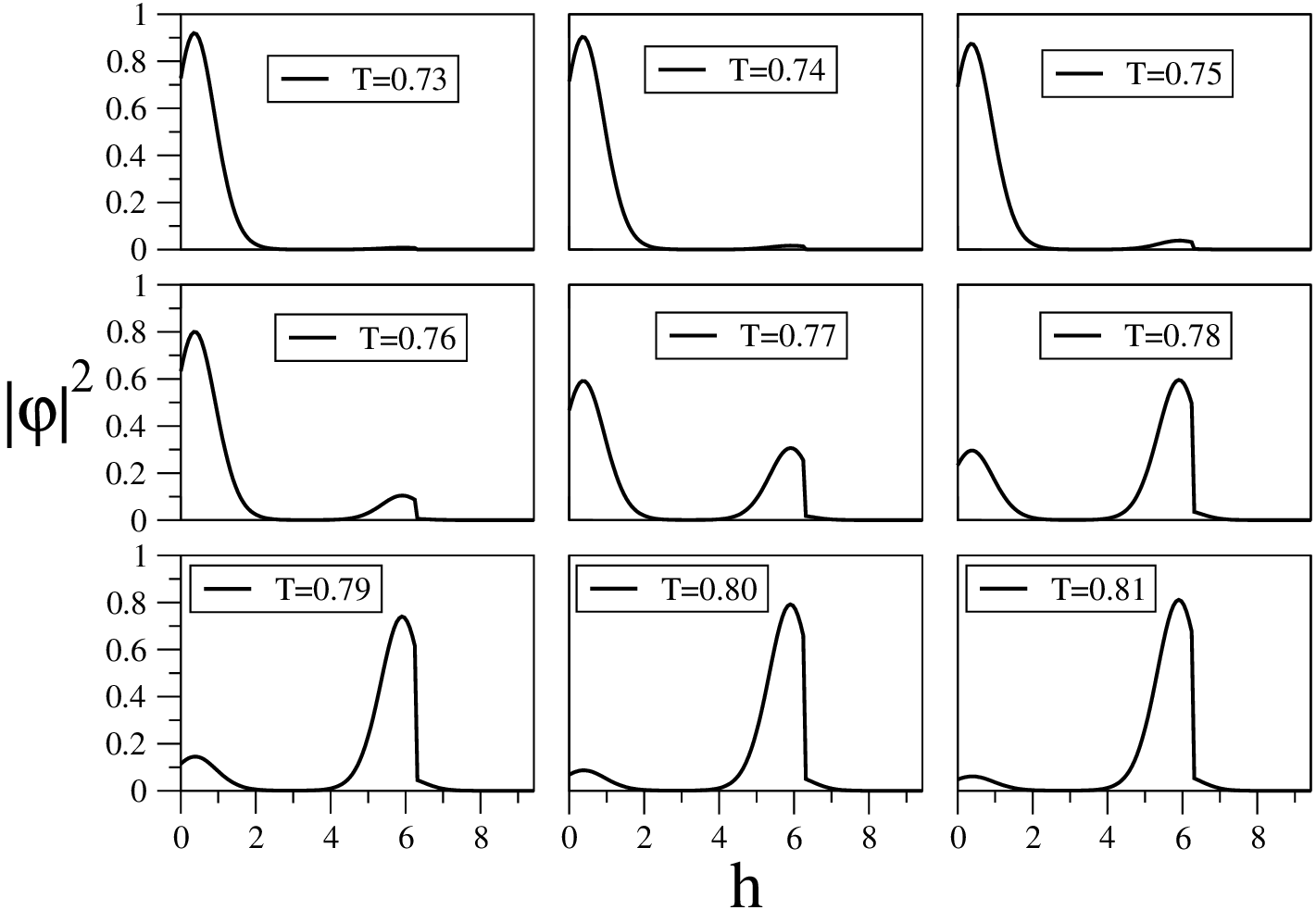}
\caption{\label{fig:9im}Probability density of $h$ for different temperatures
around the narrow peak of the specific heat for $M=1440$ and $h_{max}=100$
($\Delta h=5/72$).}
\end{figure}

We have also computed, using Eq.\ (\ref{eq:specific}), the specific heat from the
numerically obtained eigenvalue. This is shown in Fig.\ \ref{fig:shtrans}. The jump of
\noindent
the specific heat at $T\approx 10.3$ is the jump associated
with the phase transition. The peak at $T\approx 1.4$ is a well-known Schottky
anomaly (see, e.g., \cite{us} and references therein) related to the fact that
the heights pass from being mostly in one well of the cosine potential to expand
to different wells. There is an extra feature, namely the narrow peak at $T\approx 0.4$.
If we look at the gap $\Delta \epsilon$ between the two first eigenvalues, it
effectively has a minimum at that temperature, what would make us think of an additional
phase transition. Furthermore, that transition would have a physical
interpretation. In Fig.\ \ref{fig:9im} we represent the square value of the
\noindent
first eigenvalue of the transfer operator, that as we saw in Eq.\
(\ref{eq:averages}) has the interpretation of the probability density of $h_i$. In
the figure we see that at the temperature of the transition ($T\approx 0.77$ for
the parameters of the figure) the heights pass from being almost all in the
lowest $h$ well of the cosine potential
(the potential well with the minimum at $h=0$)
to being in the highest $h$ one (the well with minimum at $h=2\pi$). This
"transition", however, does not survive a finite-size study: as, keeping
$h_{max}$ fixed, we make $\Delta h$ smaller, the temperature of the transition
goes to zero, showing us that it is nothing but a result of the discretization
and the numerical technique employed. Our Monte Carlo simulations will confirm
this, as they show all the way down to the lowest temperature we have simulated,
below $T=0.1$, that the well preferred by the heights is the highest $h$ one
(see Sec.\ \ref{sec:MC} and Fig.\ \ref{fig:well} below). Upon this observation,
one question immediately
arises: if both the first and the second well of the cosine
potential are energetically equally favorable, why does the system choose as the
equilibrium one the second? The reason is that entropically they are not the
same, and the configuration of the heights in the highest $h$ well has greater
entropy. The reason for this is that the only escape a height $h_i$ has
from the lowest $h$ well is going to the highest $h$ one (at low
enough temperatures at which big $h$ differences are very unlikely).
But from the highest $h$
well it can escape to the lowest $h$ one, or to the next cosine well outside
the Burkhardt-like square well. So the two wells are not symmetrical, and the
configurations in the highest $h$ one have higher entropy.
In that way, what we see in the lowest temperature curves in Fig.\
\ref{fig:9im} would be in fact a metastable state with higher free energy than
the true equilibrium one, the heights in the highest $h$ well.

\section{Monte Carlo simulations}
\label{sec:MC}

To confirm the conclusions drawn from the analytical simulations on the
existence of a phase transition we have resorted to
parallel tempering Monte Carlo simulations \cite{us3,newman,iba}.
Representative configurations at a given temperature are generated with
a heat bath algorithm \cite{toral,us3}, 
in which new values $h'_i$ for the height at site $i$ are proposed
according to the rule
\begin{equation}
\label{toral1}
h_{i}'=\frac{h_{i-1}+h_{i+1}}{2}+\xi \sqrt{\frac{T}{2J}},
\end{equation}
$\xi$ being a Gaussian random variable of zero mean and unit
variance, and are accepted with a probability
$\min[1,{\rm e}^{-\delta\mathcal{H}/T}]$ with $\delta{\mathcal H}=[V(h_
{i}')-V(h_{i})].$ The reason to accept or reject using only the potential
term in the Hamiltonian is that the proposal in Eq.\ \ref{toral1} \emph{exactly}
reproduces the quadratic coupling fluctuations, which are gaussian. Since that
term is already fully included in the proposal, we do not need it in the
acceptance rate.

The parallel tempering algorithm
then considers simultaneous copies of the system at different
temperatures,
allowing exchange of configurations between them.
This is particularly efficient for low
temperature configurations, which are most susceptible to being trapped
in metastable regions.
The simulation
starts using a single system copy (replica) 
at the highest temperature of interest.
After simulating it we get the temperature for the next replica from
the energy fluctuations. We repeat the same process until we have a set of
temperatures that covers the whole range of interest. Then we run a parallel
tempering simulation of all replicas and from it get improved values of the
temperature set. This auto-tuning process continues until we have an almost
perfect measure of the specific heat, which shows that we are using a near to
optimal temperature set, and at the same time that the different replicas
are properly equilibrated. After allowing this last temperature set replicas
run for further equilibration, we start the measuring run.
\noindent
\begin{figure}
\vspace*{2mm}
\includegraphics[width=8.0cm]{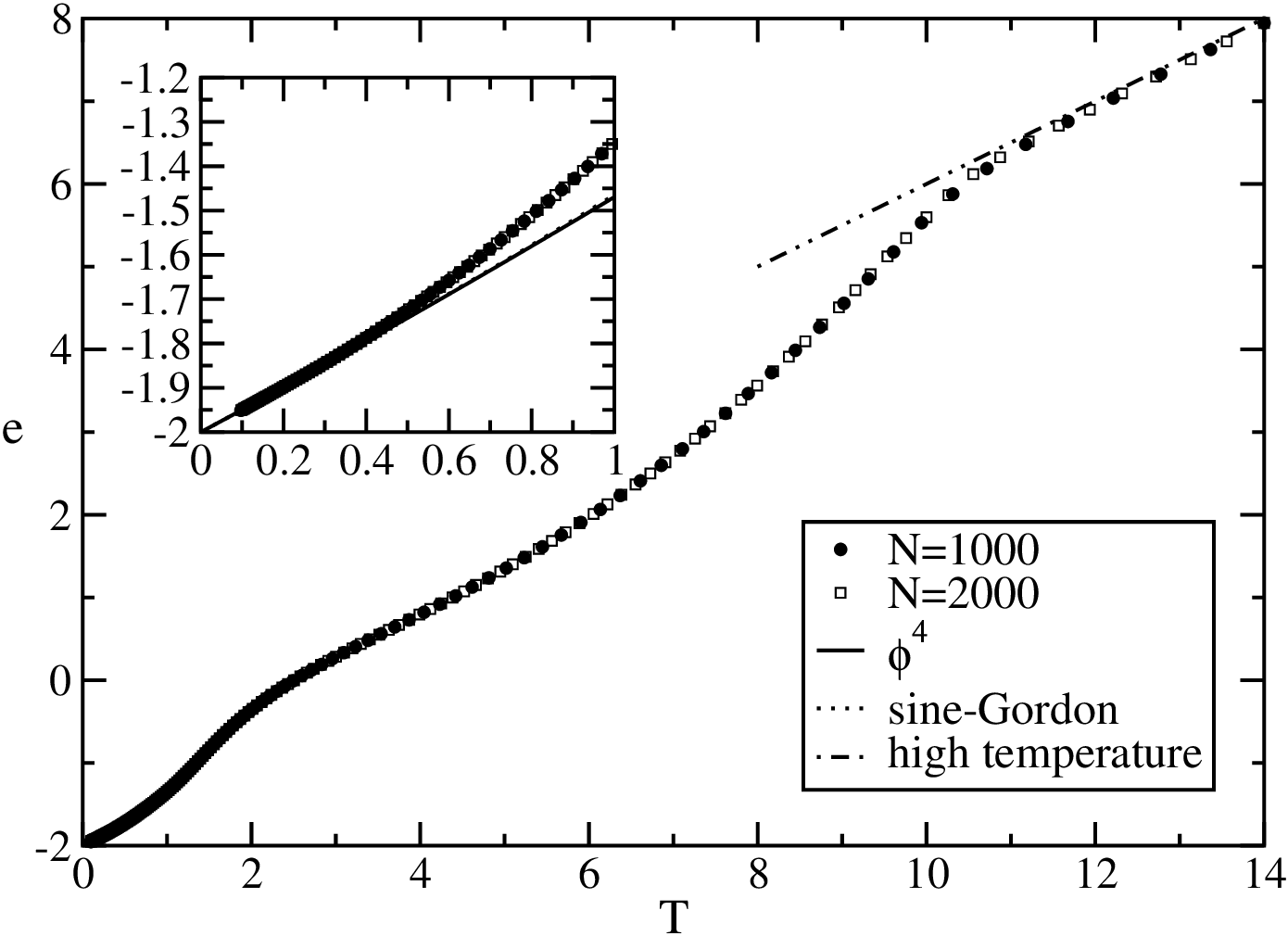}
\caption{\label{fig:energy}Internal energy per particle
obtained from Monte Carlo
simulations. Inset: display of the low temperature region and comparison with
the predictions of Eqs.\ (\ref{eq:e_phi4}) and (\ref{eq:e_sG}). Note that the zero
temperature energy is shifted by $-2$ with respect to Eqs.\ (\ref{eq:e_phi4}) and
(\ref{eq:e_sG}) to take into account the square well potential.
Lines are as indicated in the plot.  At this scale, the predictions of the
sine-Gordon and the $\phi^4$ models (in the inset) are indistinguishable.}
\end{figure}
\noindent
\begin{figure}
\vspace*{2mm}
\includegraphics[width=8.0cm]{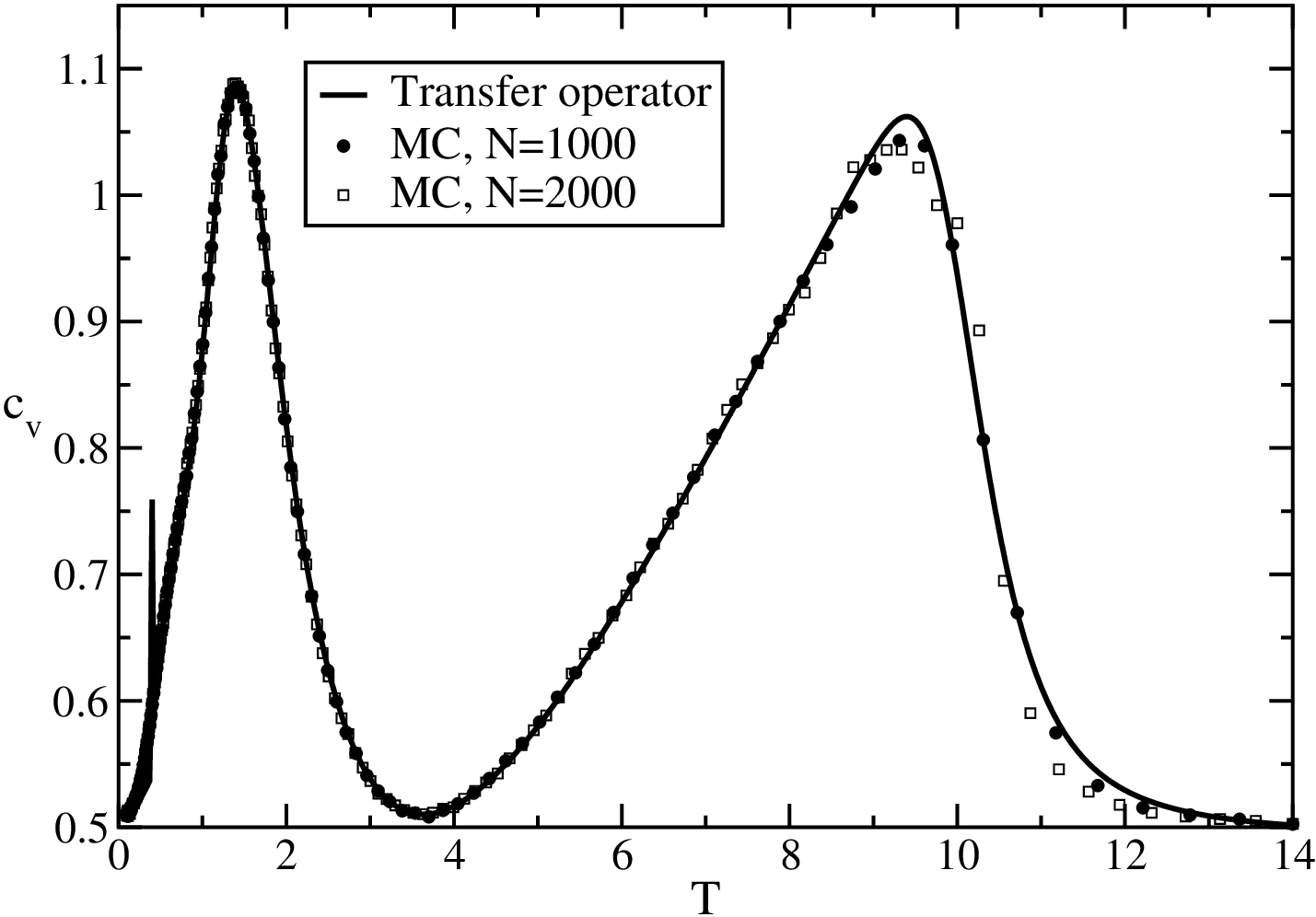}
\caption{\label{fig:shMC} Specific heat from Monte Carlo simulations, comparison
is made with the numerical transfer operator result. Error bars of the
simulations are of the size of the symbols or smaller. Symbols and
lines are as indicated in the plot.}
\end{figure}
\noindent
\begin{figure}
\vspace*{2mm}
\includegraphics[width=8.0cm]{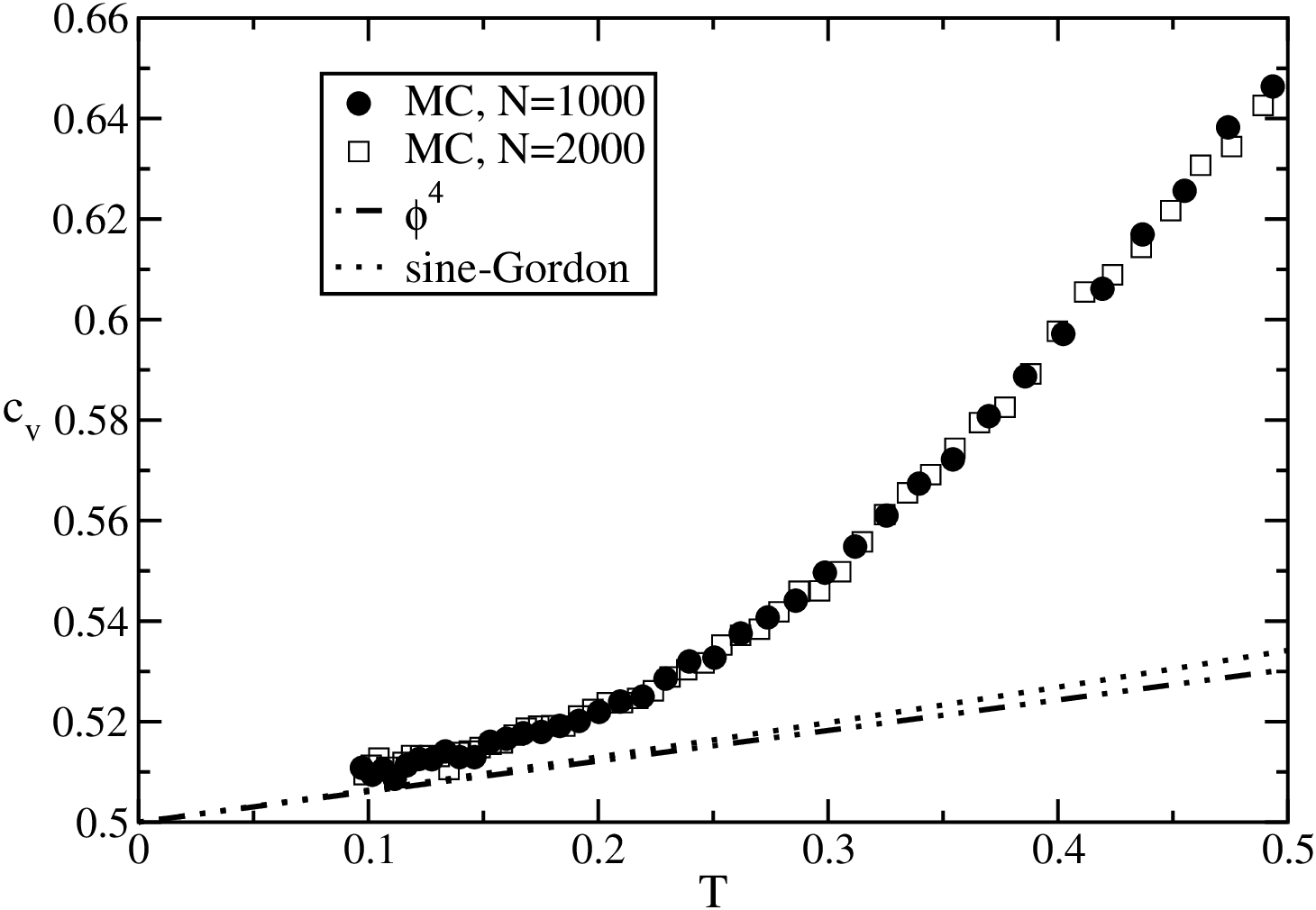}
\caption{\label{fig:Tlow}Specific heat from Monte Carlo simulations at low
temperatures compared with the predictions of Eqs.\ (\ref{eq:sh_phi4}) and
(\ref{eq:sh_sG}).
The symbols are simulation results for different system sizes as
indicated in the plot. Error bars are of the size of the symbols.
The dashed-double dotted line
is the prediction of Eq.\ (\ref{eq:sh_phi4})
and the dotted one the prediction of Eq.\ (\ref{eq:sh_sG}).}
\end{figure}

The parameters we have used for our simulations, as already said, are $V_0=1$,
$U_0=2$ and $R=2\pi$. We have also ran simulations with different values of the
parameters without finding qualitative differences. We have made simulations for
system sizes of $N=500$, $N=1000$ and $N=2000$, although for simplicity we do
not present results for $N=500$. In Fig.\ \ref{fig:energy} we plot the internal
energy per particle.
We see that the results for both system sizes agree perfectly, and that
the agreement with the theoretical predictions for low temperature [Eqs.\
(\ref{eq:e_phi4}) and (\ref{eq:e_sG})] is quite remarkable. At high temperature it
has the predicted slope $0.5$, and we see at $T\simeq 10$ the change in the
slope indicating the temperature of the phase transition. Fig.\ \ref{fig:shMC}
shows the specific heat obtained from the simulations. We see that the
coincidence between both system sizes and the numerical transfer operator result
is perfect, except in the low temperature region where we have seen that the
numerical transfer operator introduces the spurious transition,
and in the region of the phase
transition, where small discrepancies due to finite size effects arise. As
should be expected, the transition is more abrupt for the largest system size,
$N=2000$. This
agreement between the results of two completely different approaches as the
numerical transfer operator and the Monte Carlo simulations provides firm grounds to our
claims. In Fig.\ \ref{fig:Tlow} we see how the specific heat has an asymptotic
behavior as $T\to 0$ in agreement with approximations (\ref{eq:sh_phi4}) and
(\ref{eq:sh_sG}).

\noindent
\begin{figure}
\vspace*{2mm}
\includegraphics[width=8.0cm]{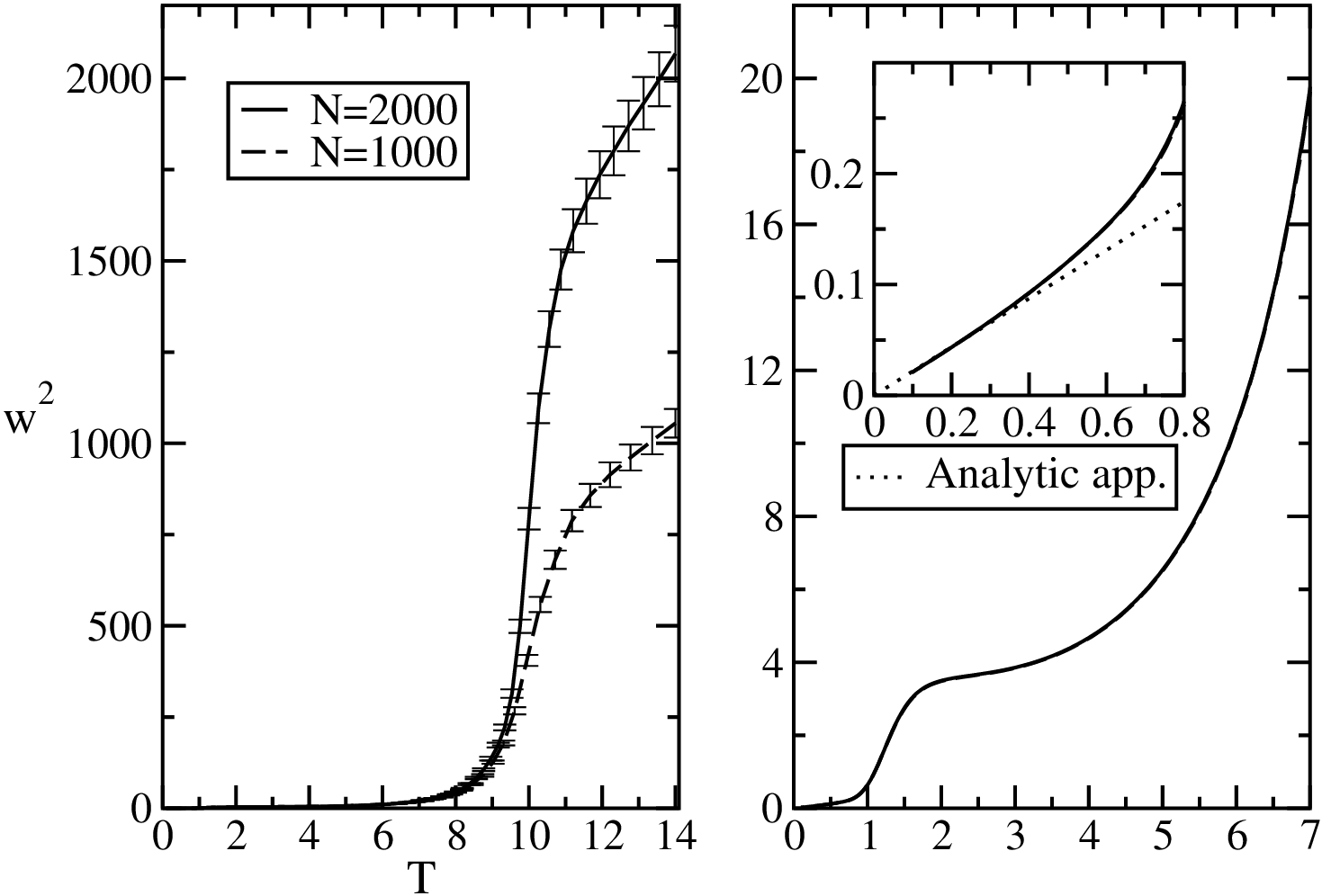}
\caption{\label{fig:rugos}Left: squared roughness $w^2$ vs.\ $T$. Right:
zoom of a lower temperature region. Note the perfect overlap of the results
for the two different system sizes below the transition temperature. Inset: yet
another zoom of an even lower temperature region, where we can see the comparison
between simulation results and the prediction of Eq.\ (\ref{eq:ruglow}).}
\end{figure}
\noindent
\begin{figure}
\vspace*{2mm}
\includegraphics[width=8.0cm]{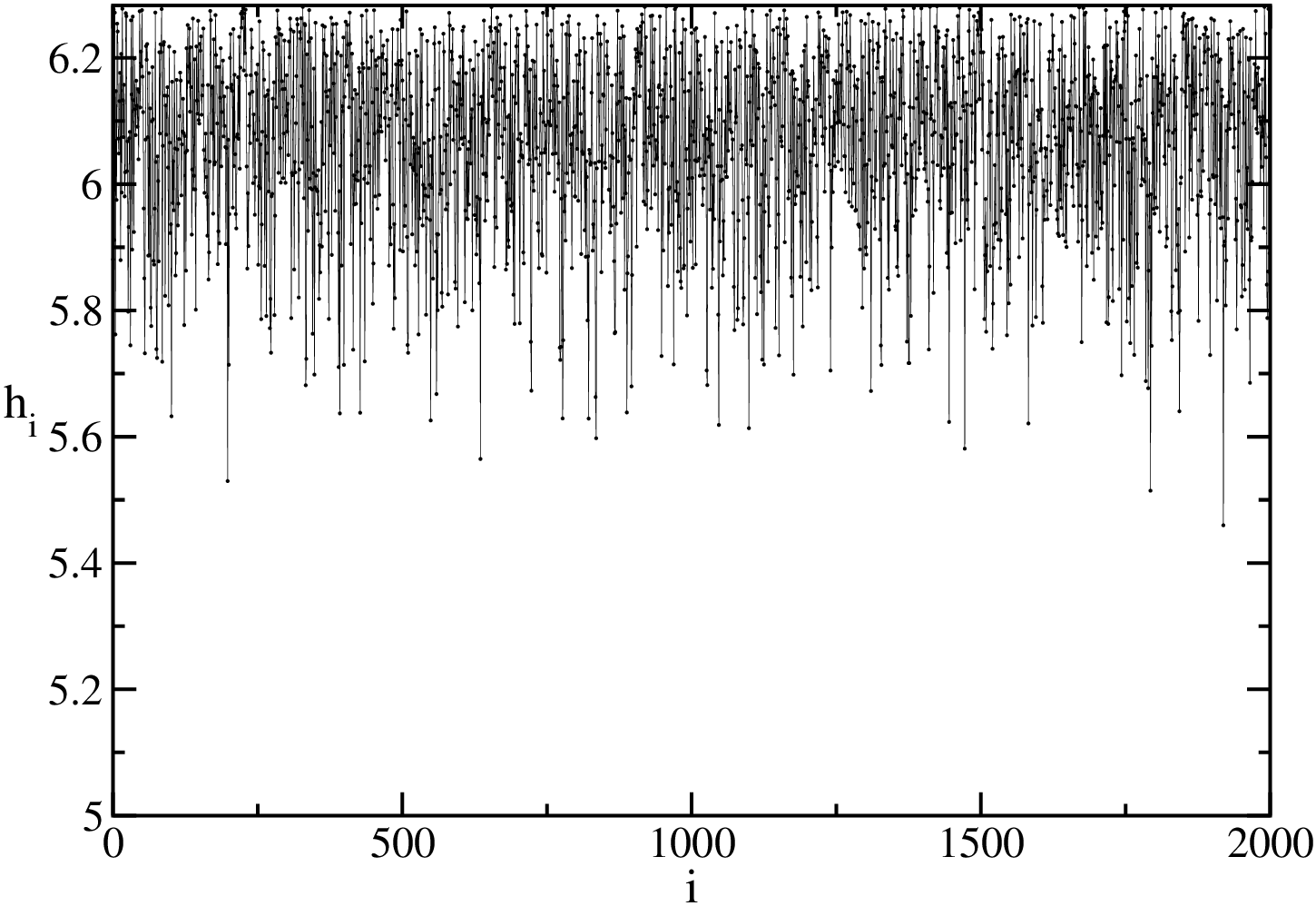}
\caption{\label{fig:well}Typical interface configuration at low temperatures.
This one is for the $N=2000$ Monte Carlo simulation at $T=0.0981$.}
\end{figure}
\noindent
\begin{figure}
\vspace*{2mm}
\includegraphics[width=8.0cm]{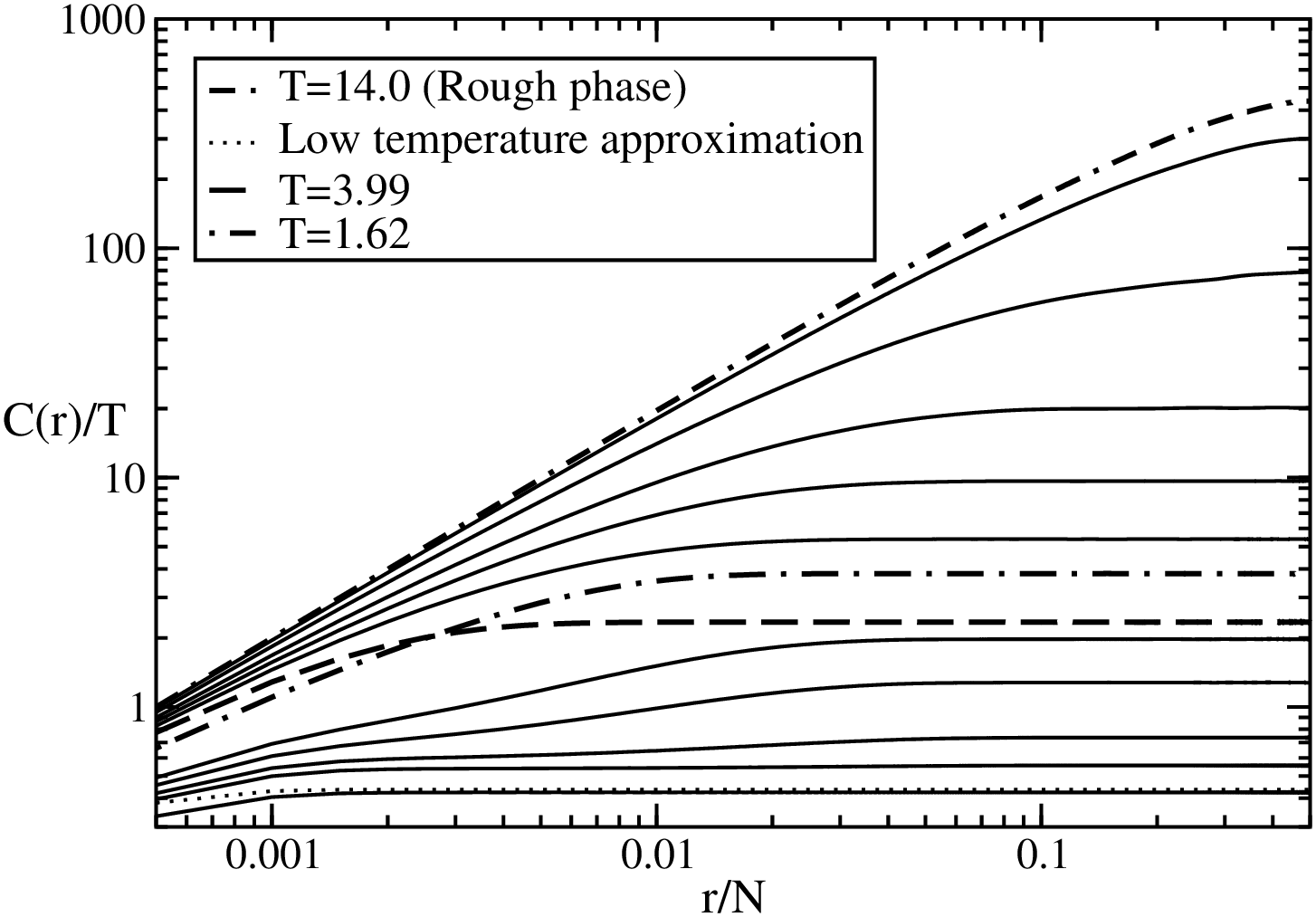}
\caption{\label{fig:correl}Height difference correlation functions scaled by the
temperature from the $N=2000$ simulation. Temperatures are (from up to down of
the greatest value:
$T=$14.0, 10.26, 9.53, 8.56, 7.80, 6.90, 1.62, 3.99, 1.12, 0.995, 0.836, 0.697,
0.0981.}
\end{figure}

As the most important verification of the transition,
Fig.\ \ref{fig:rugos} shows the squared roughness. For temperatures above the
phase transition, $w^2$ becomes dependent of the system size and diverges
with $N$, showing us that we are in a rough phase. Below $T_c$, the
results for both system sizes are the same, and as $T\to 0$ we see the behavior
predicted in Eq.\ (\ref{eq:ruglow}). The step in the roughness between $T\simeq
1$ and $T\simeq 1.5$ is an effect of the Schottky anomaly \cite{us3} we have already
mentioned.
Between
$T\simeq 2$ and $T\simeq 4$, we see a little plateau in the roughness curve.
This plateau is caused by the dominating part that the kinks formed between
the lowest $h$ well and the highest $h$ one play at these temperatures,
while the relaxation of the
heights in each well as temperature goes down is almost screened by the
effect of the kinks in the roughness. At the lowest temperatures, below $T\simeq 1$,
all effects of kinks disappear, and the interface is trapped in the highest $h$ well
in the square potential, as we already noted above and show in Fig.\ \ref{fig:well}. 
This is related to the apparent phase transition studied in \cite{us3}.

In Fig.\ \ref{fig:correl} we see the height-difference correlation function,
scaled by temperature, from the simulations with $N=2000$. All the curves
corresponding to temperatures higher than $T_c$ collapse to a single curve.
This is the expected behavior for the high temperature rough phase,
as the potential term in the Hamiltonian is expected to be irrelevant
at these temperatures, leaving us only with the quadratic coupling, which is
the Edwards-Wilkinson model \cite{EW} that predicts exactly this independence
of $T$ for $C(r)/T$, see also \cite{us}.
The
first curve below this collapse is the curve for $T=10.26$. So, from our simulations
we obtain $T_c=10.26$, in excellent agreement with the numerical transfer
operator result. For $N=1000$ (not shown) we obtain $T_c=10.31$ and the same behavior
depicted in Fig.\ \ref{fig:correl}. Finally, note the agreement between the
prediction of Eq.\ (\ref{eq:corrlow}) and the actual low temperature correlation
functions we find in simulations.
We see again in Fig.\ \ref{fig:correl} the effect of kinks that appeared
in the roughness between $T\simeq 2$ and $T\simeq 4$: the temperature scaled
height-difference correlation function has a non monotonous behavior with
temperature between $T=3.99$ and $T=1.62$. In this range the different functions
(without scaling)
are almost independent of temperature, so the scaled functions have higher values
as we reduce temperature. Note that this behavior only appears above certain
length scale. At very short scales, the effect of kinks has little importance
(as we need a certain system size to have probabilities of kinks to appear) and the
relaxation of heights continues with decreasing temperature.


\section{Conclusions}
\label{sec:conc}

We have studied in detail a model first proposed by us \cite{super},
which combines the model proposed by
Burkhardt in \cite{burkhardt} and the well known sine-Gordon model. We
show here by analytic approximations and by
two different numerical methods (transfer operator and
Monte Carlo simulation) that it has a continuous phase transition between a high
temperature rough phase and a low temperature flat one. We have characterized the
thermodynamics of the model, establishing  its non trivial behavior in the
flat phase due to interaction of the two kinds of forces (periodic potential and
substrate attraction) present in it. This gives rise to the existence of a
temperature region (between $T\simeq 1.6$ and $T\simeq 4.0$) where physical
magnitudes of the interface as roughness and spatial correlations are
quite independent of the temperature. In addition, our work also stands as
a careful study of a 1D thermodynamical phase transition. While we hope our 
results will stimulate further 
studies in this field, misunderstood for a long time,
we want to add a few caveats about how
numerical results can
lead to misleading conclusions. First, we have seen that 
the numerical analysis of the transfer operator produced an artifact which 
looked like a second phase transition in the low temperature regime. 
Second, we have shown in a previous paper \cite{us3} that simulations 
can yield results reminiscent of a true phase transition even for 
extremely large system sizes, whereas it is rigorously known \cite{us2}
that such transition is impossible.  Therefore, it must be born in mind
that only a judicious 
combination of theoretical results, numerical analysis and simulations
may provide firm grounds to claims of existence of phase transitions in 
models that are not exactly solvable. This is even more important in the
case of 1D systems, where the debate is contaminated by the false 
prejudices against their own existence \cite{us4}.

Finally, we want to stress that 
the results we have obtained on this model
suggest a more amenable analytical
and computationally way to study the properties of
modified versions of the
2D sine-Gordon model as we did in \cite{super}
for the random substrate version.
As our model has a transition
between a low temperature flat phase and a high temperature rough one, just like
the 2D sine-Gordon model without disorder,
in that work we showed how the
addition of disorder to our model can give us insight of what happens in the low
temperature phase of the 2D random sine-Gordon model. We believe that the same 1D
approach to 2D problems will prove fruitful in many other contexts. Its two main
advantages are that usually 1D models are more amenable to analytical treatment
than 2D ones, and that simulating a 1D model requires much less computational
effort. We hope that many new insights will be obtained following this line 
of research.

\begin{acknowledgments}
This work has been
supported by the Ministerio de Ciencia y Tecnolog\'\i a of Spain
through grant BFM2003-07749-C05-01.
\end{acknowledgments}


\begin{thebibliography}{88}
\bibitem{Plischke} M.\ Plischke and B.\ Bergersen, {\em Equilibrium Statistical Physics}
(World Scientific, Singapore, 1994), Chap.\ 11 and references therein.
\bibitem{Weeks} J.\ D.\ Weeks and G.\ H.\ Gilmer, Adv.\ Chem.\ Phys.\
{\bf 40}, 157 (1979); J.\ D.\ Weeks, in {\em Ordering in Strongly Fluctuating
Condensed Matter Systems}, edited by T.\ Riste (Plenum, New York, 1980);
\bibitem{Beijeren}
H.\ van Beijeren and I.\ Nolden, in {\em Structure and
Dynamics of Surfaces}, edited by W.\ Schommers and P.\ von Blackenhagen,
Topics in Current Physics vol.\ 43
(Springer, Berlin, 1987).
\bibitem{Barabasi} A. L. Barab\'asi and H.\ E.\ Stanley, {\em Fractal
Concepts in Surface Growth} (Cambridge University Press, Cambridge, 1995).
\bibitem{Krug} J.\ Krug, Adv.\ Phys.\ {\bf 46}, 139 (1997).
\bibitem{Pimpinelli} A.\ Pimpinelli and J.\ Villain, {\em Physics of Cristal Growth}
(Cambridge University Press, Cambridge, 1998).
\bibitem{Toner} J.\ Toner and D.\ P.\ DiVincenzo, Phys.\ Rev.\ B {\bf 41},
632 (1990).
\bibitem{Shapir} Y.\ Shapir, in {\em Dynamics of fluctuating interfaces
and related phenomena}, D.\ Kim, H.\ Park, B.\ Kahng, eds.\ (World
Scientific, Singapore, 1997).
\bibitem{us} A. \ S\'anchez, A.\ R.\ Bishop and E.\ Moro, Phys.\ Rev.\
E {\bf 62}, 3219 (2000).
\bibitem{Riegercom} H.\ Rieger, Phys.\ Rev.\ Lett.\  {\bf 74}, 4964 (1995).
\bibitem{Marinari} E.\ Marinari, R.\ Monasson, and J.\ J.\ Ruiz-Lorenzo,
J.\ Phys.\ A {\bf 28}, 3975 (1995).
\bibitem{Lancaster} D.\ J.\ Lancaster and J.\ J.\ Ruiz-Lorenzo, J.\ Phys.\
A {\bf 28}, L577 (1995).
\bibitem{Zeng} C.\ Zeng, A.\ A.\ Middleton, and Y.\ Shapir, Phys.\ Rev.\
Lett.\ {\bf 77}, 3204 (1996).
\bibitem{Rieger1}
U.\ Blasum, W.\ Hochst\"attler, U.\ Moll, and
H.\ Rieger, J.\ Phys.\ A {\bf 29}, L459 (1996).
H.\ Rieger and U.\ Blasum, Phys.\ Rev.\ B {\bf 55},
R7394 (1997).
\bibitem{Coluzzi} B.\ Coluzzi, E.\ Marinari, and J.\ J.\ Ruiz-Lorenzo,
J.\ Phys.\ A {\bf 30}, 3771 (1997).
\bibitem{us1} A.\ S\'anchez, A.\ R.\ Bishop, D.\ Cai,
N.\ Gr\o nbech-Jensen,
and F.\ Dom\'\i nguez-Adame, Physica D {\bf 107}, 325 (1997).
\bibitem{jjfises} J.\ J.\ Ruiz-Lorenzo, in {\em Proceedings of the
VIII Spanish Meeting on Statistical Physics FISES '97}, edited by
J.\ A.\ Cuesta and A.\ S\'anchez (Editorial del Ciemat, Madrid, 1998).
\bibitem{Rieger2} H.\ Rieger, Phys.\ Rev.\ Lett.\ {\bf 81}, 4488 (1998).
\bibitem{super} S.\ Ares, A.\ S\'anchez and A.\ R.\ Bishop, Europhys.\ Lett.\ {\bf 66},
552 (2004).
\bibitem{vanHove} L.\ van Hove,
Physica \textbf{16}, 137 (1950) (reprinted in  \cite{lieb}, p.\ 28)
\bibitem{lieb}
E.\ H.\ Lieb and D.\ C.\ Mattis, eds., {\em Mathematical Physics
in One Dimension} (Academic, New York, 1966).
\bibitem{landau} L.\ D.\ Landau and E.\ M.\ Lifshitz, {\em Statistical
Physics Part 1} (Pergamon, New York, 1980).
\bibitem{us4} J.\ A.\ Cuesta and A.\ S\'anchez, J.\ Stat.\ Phys.\ {\bf 115}, 869 (2004).
\bibitem{forgacs} G.\ Forgacs, J.\ M.\ Luck, Th.\ M.\ Nieuwenhuizen and H.\ Orland,
Phys.\ Rev.\ Lett.\ {\bf 57}, 2184 (1986).
\bibitem{derrida} B.\ Derrida, V.\ Hakim and J.\ Vannimenus, J.\ Stat.\ Phys.\ {\bf 66},
1189 (1992).
\bibitem{burkhardt} T.\ W.\ Burkhardt, J.\ Phys.\ A {\bf 14}, L63 (1981).
\bibitem{chui} S.\ T.\ Chui and J.\ D.\ Weeks, Phys.\ Rev.\ B {\bf 23},
R2438 (1981).
\bibitem{leeuwen} J.\ M.\ J.\ van Leeuwen and H.\ J.\ Hilhorst,
Physica (Amsterdam) {\bf 107A}, 319 (1981).
\bibitem{schiff} L.\ I.\ Schiff, {\em Quantum Mechanics} $2^{nd}$ edn.
(McGraw-Hill, New York, 1968).
\bibitem{us2} J.\ A.\ Cuesta and A.\ S\'anchez, J.\ Phys.\ A {\bf 35}, 2377
(2002).
\bibitem{us3} S.\ Ares, J.\ A.\ Cuesta, A.\ S\'anchez and R.\ Toral,
Phys.\ Rev.\ E {\bf 67}, 046108 (2003).
\bibitem{toral} R.\ Toral, in {\em Proceedings of the Third Granada
Lectures in Computational Physics}, edited by P.\ L.\ Garrido and J.\ Marro,
(Lecture Notes in Physics vol.\ 448, Springer-Verlag, Berlin-Heidelberg, 1994).
\bibitem{schneider} T.\ Schneider and E.\ Stoll, Phys.\ Rev.\ B {\bf 22}, 5317
(1980).
\bibitem{ch} R.\ Courant and D.\ Hilbert, {\em Methods of Mathematical Physics
vol.\ I} (Wiley Classics, New York, 1989).
\bibitem{EW} S.\ F.\ Edwards and D.\ R.\ Wilkinson, Proc.\ R.\ Soc. London,
Ser.\ A {\bf 381}, 17 (1982).
\bibitem{theo} N.\ Theodorakopoulos, Phys.\ Rev.\ E {\bf 68}, 026109 (2003).
\bibitem{dp} T.\ Dauxois and M.\ Peyrard, Phys.\ Rev.\ E {\bf 51} 4027 (1995);
T.\ Dauxois, N.\ Theodorakopoulos and M.\ Peyrard, J.\ Stat.\ Phys.\ {\bf
107}, 869 (2002).
\bibitem{thierry} T.\ Dauxois, {\em Nonlinear dynamics and statistical
mechanics of a model for DNA}, Ph.\ D.\ Thesis, University of Dijon (1993). 
\bibitem{newman} M.\ E.\ J.\ Newman and G.\ T.\ Barkema,
{\em Monte Carlo Methods in Statistical Physics} (Oxford University,
Oxford, 1999).
\bibitem{iba} Y.\ Iba, Int.\ J.\ Mod.\ Phys.\ {\bf 12}, 623 (2001).



\end{thebibliography}
\end{document}